# Computational workflow for investigating hydrogen permeation in novel hydrogen storage materials

Sourabh Singha[1] and Abhijit Chatterjee[1]

[1] Department of Chemical Engineering, IIT Bombay, Mumbai 400076, India

abhijit@che.iitb.ac.in

**Abstract.** The United States Department of Energy (DOE) has set ambitious targets for hydrogen storage materials for onboard light-duty cars which are to be achieved by 2027. One of the major problems in solid hydrogen storage materials is the sluggish uptake/release kinetics. Much attention has been focused on understanding kinetics. Hydrogen solid-state diffusion is a rate-controlling step in the majority of metal hydrides. Here we discuss computational workflow which can be used to estimate hydrogen diffusivity. A detailed study of hydrogen concentration, hydrogen neighbors is performed on nickel hydride (NiH) fcc materials to understand their effect on H diffusion. The nudged elastic band (NEB) method is used to determine hydrogen diffusion barrier with various hydrogen concentrations in presence of hydrogen neighbors. The energy barriers for hydrogen hopping were calculated for few million different configurations with various local chemical environments. Two paths for H hopping from one octahedral site to a vacant neighbor octahedral site are identified: one path is straight, and the other is curved via tetrahedral site. The curved path shows diffusion faster than the straight path. This study demonstrates H diffusion is faster at higher hydrogen concentration, as the concomitant volume expansion lowers the energy barrier.

**Keywords:** Activation energy, Nudged elastic band, Kinetics, Nickel hydride, Hydrogen storage, Diffusion.

# 1 Introduction

Hydrogen, a potential future fuel, offers high energy density and clean energy in fuel cells [1]. However, one of the most pressing issues in the use of hydrogen as a fuel is the need for safe and effective storage particularly for electric vehicle applications. For instance, storing hydrogen in gaseous and liquid states has safety issues because of the high pressure and extremely low temperatures [2]. A possible solution is to store hydrogen in metals (e.g., metal hydrides), which can provide high capacity and minimize accidental risk; however, the overall capacity is limited by sluggish kinetics and high ad/absorption and desorption temperatures [3]. The U.S. DOE has set a target (gravimetric capacity 6.5 wt.%, sorption/desorption temperature -40/80°C, cycle life 1500, and hydrogen release and storage in 3-5 minutes) for hydrogen storage using various materials, including metal hydrides and carbon-based materials like nanotubes and graphene [4–6]. However, these materials face challenges such as poor kinetics and high adsorption and desorption temperatures.

According to Zhou et al. [7], solid-state diffusion is typically the rate-limiting step for hydrogen uptake and release. Examples include Pd [8] and Mg [9] hydrides, in which diffusion is the rate-controlling step. Therefore, in most metal hydrides, studying diffusion behavior during hydrogen uptake and release becomes essential [10]. Ni-based hydride are promising materials being explored [11]. For this reason, Ni hydride is studied here to establish a computational protocol. Hydrogen diffusion is thermally activated. Thus, accurate calculation of activation energy becomes important. The rate constant for hydrogen hopping within the solid material obeys the Arrhenius equation

$$k = k_0 \exp\left(-\frac{E_a}{k_B T}\right). \quad (1)$$

In equation (1), k is rate constant, $k_0$ is pre-factor (typically $k_0 = 10^{12}\ s^{-1}$), $E_a$ is activation energy, T is temperature and $k_B$ is Boltzmann constant ($k_B = 8.617 \times 10^{-5}$ eV/K). The H hopping rates likely depend on a number of factors, including the overall hydrogen concentration, hydrogen neighbors and the arrangements. However, such information is not available in literature. To address this gap, the NEB method has been used here to calculate hydrogen hopping barrier for a large number of configurations with a focus on Ni hydride. Our study shows H diffusion becomes faster as the hydrogen concentration increases and the reason for this is the significant volume expansion (19.94%) that occurs during the first-order α to β phase transition. This volume expansion opens up the structure, making it easier for the hydrogen atom to hop which results in a reduction in energy barrier. In case of hydrogen neighbors, the bridge position neighbors promote higher energy barrier as the H atom faces difficulty while passing bridge position while hopping from one octahedral site to another.

## 2  Simulation methods

### 2.1  Nudged elastic band calculation

The energy barrier calculation is performed using Large-scale Atomic/Molecular Massively Parallel Simulator (LAMMPS) code [12] using an automated computational workflow which includes codes/script in Fortran90, bash script and MATLAB. All simulations involve periodic boundary conditions. The system is bulk NiH (100) fcc cubic, and lattice parameters are a=b=c=3.52 Å [13] and angle α=β=γ=90° with space group Fm$\bar{3}$m (No.225). The NiAlH Embedded Atom Method (EAM) potential [14] has been used for NiH calculation as EAM potential is good fit for metal system such as Ni, Pd [13][15]. The fcc structure contains both tetrahedral and octahedral sites that can capture hydrogen. Our studies focus on H at the octahedral sites, as they are known to be more favorable than tetrahedral sites [16], and our calculations support this observation. Therefore, H atoms are allowed to occupy these sites.

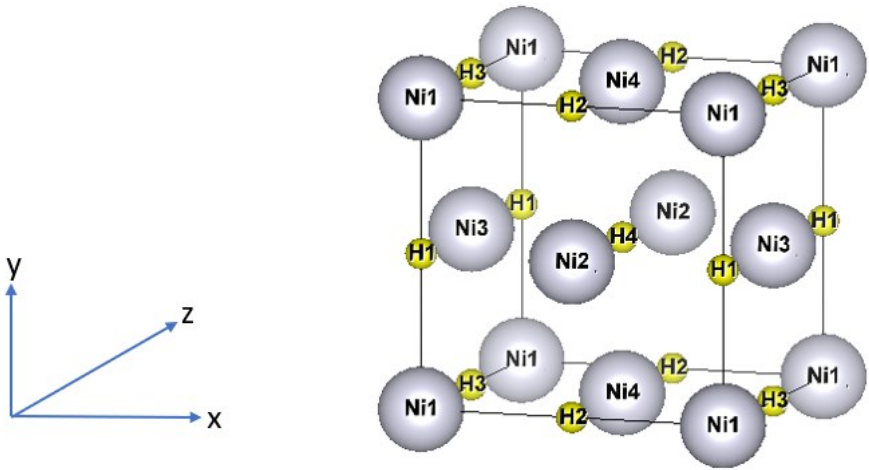

Fig. 1: Basic unit cell of (a) NiH (contains 4 Ni and 4 H atoms).

For NEB calculations, a system size with 10×10×10 unit cells has been chosen. The NEB calculation requires stable initial and final structures. LAMMPS software was used for the NEB simulations, with nine intermediate images created along the reaction pathway. The simulation was performed with a force tolerance of $10^{-2}$ eV/Å and a time step of 5 fs for quickmin optimization. Quickmin is an optimization algorithm that minimizes the energy of each NEB image while maintaining as straight a path as possible, which is computationally fast and inexpensive compare to other minimization methods [17]. When compared to molecular dynamics [18–21], the nudged elastic band method [22–24] is orders of magnitude faster.

## 2.2 Hydrogen hopping

Hydrogen hopping is a thermally activated event. As a consequence, hydrogen diffusion becomes dependent on temperature, such that diffusion becomes faster at higher temperature. The NEB calculation is a static calculation and does not contain temperature as a parameter while calculating energy barrier. The advantage is that rate kinetics can be calculated at different temperatures using the Arrhenius equation. However, there can be distribution of activation barriers depending on the local chemical environment. To probe this aspect, different hydrogen arrangements in the first nearest neighbor (1NN) position are considered to determine the energy barrier of the NiH system. The H atom in $NiH_x$ can have utmost 12 H neighbors. Figure 2(a) shows the 12 neighbors of the hopping H at the initial site and the 12 neighbors when H atom is at the final position.

When the H atom hops, there are 18 neighbors in the 1NN position as can be seen in Fig. 2(a). For easier visualization in Fig. 2(b) shows clearly how H hops from site 1 (one octahedral site) to site 2 (another octahedral site). The H hopping path follows a straight line if bridge neighbors are present, which leads to a high-energy barrier. Alternatively, another path, if available, taking the hydrogen atom through the tetrahedral site results in a low-energy barrier if bridge neighbors are not present is shown in Fig. 2 (b).

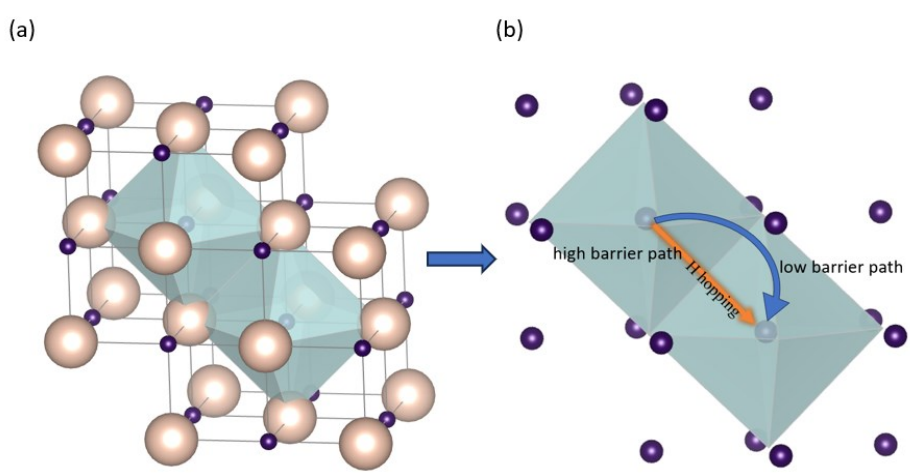

Fig. 2: (a) Structure of NiH system with 18 neighbors, (b) Hopping of H atom is (orange color) occurred one octahedral site to another. Curve path for low energy barrier and straight-line path for high energy barrier.

Considering all possible arrangements in 1NN position, around 0.262 million ($2^{18}$=262144) are considered. The NEB method finds the minimum energy path (MEP) by locating the saddle point, which corresponds to the highest energy configuration along the pathway. The energy barrier was recorded using a MATLAB code each time after the NEB calculation. A wide distribution of energy barriers is observed when hydrogen hops from one octahedral site to another, indicating the importance of neighbors. For detailed study, the NEB calculation is performed for 1NN with different H concentrations (x=0.001, 0.25, 0.50, 0.75, and 0.99) to see the energy behavior at different concentrations.

## 3   Numerical results and discussion

Initially, the activation energies calculated without any H neighbors and the activation energy is 0.57 eV as shown in Fig. 3(a). This barrier is close to the 0.45 eV barrier reported by Won et. al [25]. To illustrate the effect of the neighbors, configurations are considered in which only one neighbor site is filled with hydrogen (out of 18 neighbors, one neighbor occupied 17 neighbors unoccupied, and this process was followed for each neighbor). It is observed that there is a distribution of energy barriers as in shown Fig. 3(b). In Fig. 3(b) hydrogen 18 neighbors were mentioned by numbering from 1 to 18. The energy barrier varies from 0.45 eV to 0.61 eV while adding H neighbor in one site and 17 H neighbors were not filled at this was done at low concentration (x=0.001). This highlights the importance of neighbors site and motivates the need to investigate fully the neighbor effect for 0.26 million configurations.

(a)                                      (b)

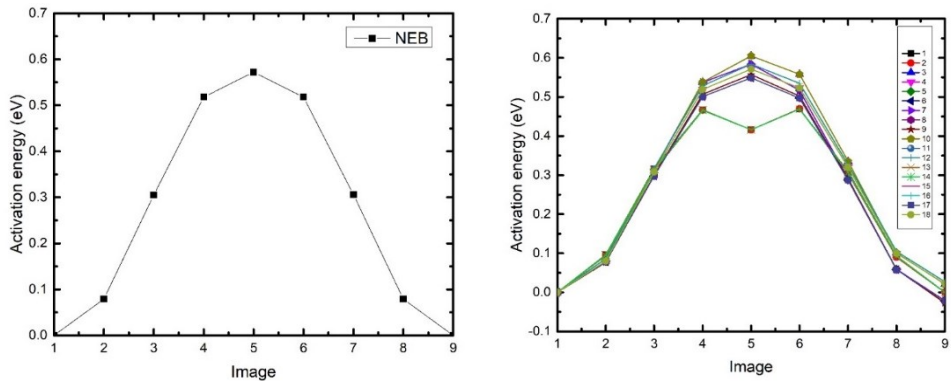

Fig. 3: Hydrogen hopping (a) from one octahedral site to a 1NN site when no H neighbors are present, (b) out of 18 neighbors, each time one H neighbor added in the NiH system and the activation energy is evaluated.

Three days were required for computing the energy barriers for 0.26 million configurations. Twenty CPUs were used to speed up the calculation. When 16-18 CPUs were used, 4-5 days were required to complete the simulation.

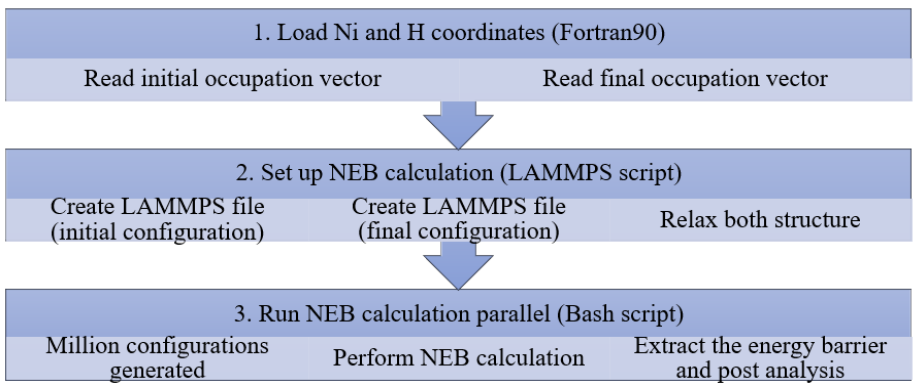

Fig. 4: Structure for the calculation of activation energy of million configuration.

Additional calculations were performed at various hydrogen concentrations (x=0.001, 0.25, 0.50, 0.75, and 0.99), and the 0.26 million energy barriers are reported for each concentration. An automated workflow was developed to run the simulations efficiently. For this, a combination of Fortran90 and bash codes was written, and a LAMMPS script was added to make the calculation parallel. The method was automated for all different H concentration calculations. The goal of the flowchart is to construct a model for activation energy barrier for million calculations and the full flowchart is described in Fig. 4.

## 3.1 H concentration

In hydrogen storage materials, the H concentration is considered to be the amount of hydrogen absorbed or stored in the metal system. Therefore, it is crucial to calculate the gravimetric and volumetric capacities.

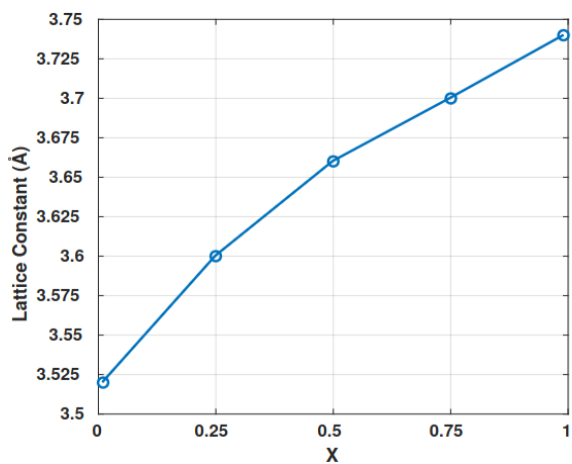

Fig. 5: The relation between lattice constant value and hydrogen concentration for NiH system.

In this study, x is the hydrogen concentration, which is defined as the ratio of the amount of H in the metal system (where x=H/Ni, H is the number of H atoms, and Ni refers to the number of Ni atoms). The goal is to estimate the energy barrier as we increase the H concentration. Five concentrations x=0.001, 0.25, 0.50, 0.75, and 0.99 are studied. While increasing the concentration to maintain the metal hydrogen distance and keeping its original structure choosing the lattice parameter is important for stable structure. Energy minimization calculations to identify the lattice constant. At x=0.001, the lattice constant value is 3.52 Å, which matches the experimental lattice constant [13]. Therefore, the same method is applied for different values of x for both NiH. Fig. 5 shows the $NiH_x$ lattice parameter as a function of x. In Fig. 5, the structure was stable at a lattice constant value of 3.6 Å for x=0.25. Similarly, lattice constant value increased as the concentration increased such as 3.66 Å, 3.7 Å, and 3.74 Å for x=0.50, 0.75, 0.99 respectively and all the case structures are stable or reach equilibrium and the lattice constant value can be seen in Fig. 5.

## 3.2  Effect of 1NN shell

At first, a low concentration (x = 0.001) was taken into consideration in order to comprehend the energy barrier caused by the neighbor effect. The result showed at 1NN during H hops on a metal surface using NEB technique, revealing an energy barrier distribution of 0.37 eV to 0.71 eV is mentioned in Fig. 6(a). The distribution illustrates that the neighbor H arrangements is important for the movement of the H atom. For further calculation other hydrogen concentrations were considered. The Ni (100) structure was filled with different hydrogen concentrations (x = 0.25, 0.50, 0.75, and 0.99). At x=0.25, the energy barrier was considerably lower than that at x=0.001, as shown in Fig. 6(b). The same trends were applicable at x=0.50, 0.75, and 0.99, where the energy barrier decreased as the hydrogen concentration increased.

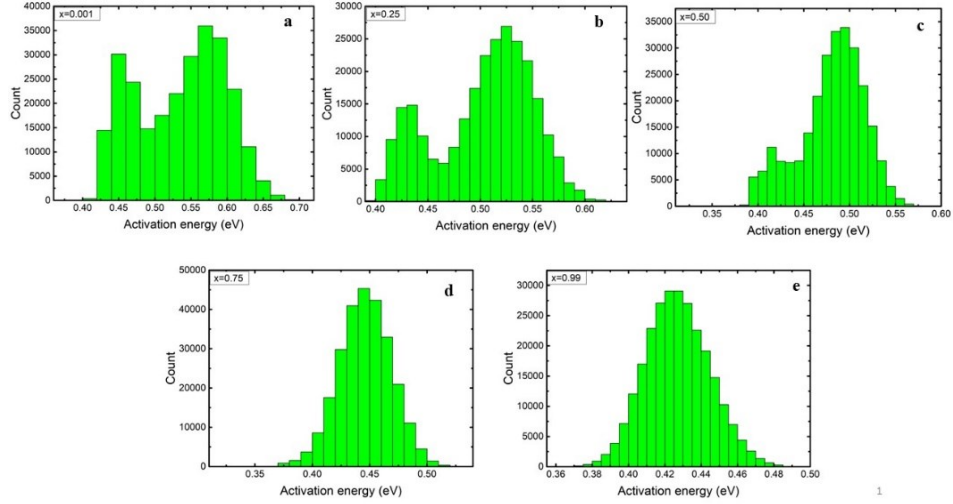

Fig. 6: The distribution of energy barrier of NiH system has shown (a) at x=0.001, (b) =0.25, (c) x=0.50, (d) x=0.75 and (e) x=0.99 (where x is the hydrogen concentration). For each case hydrogen hops from one octahedral site to another octahedral site.

As shown in Fig. 6(c)-(e), the energy barrier decreased with increasing H concentration. At x=0.99, the energy barrier range was 0.34 eV to 0.49 eV. This study can help predict the energy barrier and identify lower energy sites, which can be useful for improving the kinetics of fcc structure materials. This study will also be useful for developing accurate kinetic Monte Carlo models [26–28].

## 4 Conclusion

The effect of H concentration on the H hopping barrier is investigated while accounting for the neighbor effect. The NEB method was used to calculate the energy barrier for bulk Ni and identify low energy barrier arrangements in NiH$_x$. NEB method has been used to calculate hydrogen diffusion barrier for bulk Ni with different H concentrations (x=0.001, 0.25, 0.50, 0.75, 0.99). A wide distribution of energy barriers was noticed when H hops from one octahedral site to another, which indicates the importance of neighbors. As the H concentration increased, the energy barrier decreased because the volume expansion from the α to β phase was 19.94% for NiH while maintaining its original fcc structure, allowing hydrogen atoms to pass through. This study demonstrates that H diffusion is faster at high hydrogen concentrations because volume expansion lowers the energy barrier.